\documentclass[10pt,conference]{IEEEtran}
\IEEEoverridecommandlockouts

\usepackage[htt]{hyphenat}
\usepackage{pifont}
\usepackage{multirow}
\usepackage{graphicx}
\usepackage{booktabs}
\usepackage{subcaption}
\usepackage{float}
\usepackage{enumitem}
\usepackage{balance}

\usepackage{subcaption}
\usepackage{adjustbox}
\usepackage{caption}
\usepackage{url}
\usepackage{cite}
\usepackage{natbib}
\usepackage{algorithmic}
\usepackage{graphicx}
\usepackage{textcomp}
\usepackage{xcolor}
\usepackage{url}
\usepackage{multirow}
\usepackage[most]{tcolorbox}
\usepackage{xcolor}
\usepackage{enumitem}
\usepackage{microtype}
\usepackage[colorlinks=true, urlcolor=blue, citecolor=black, linkcolor=black]{hyperref}

\definecolor{cmarkcolor}{RGB}{21, 164, 64}
\definecolor{xmarkcolor}{RGB}{177, 0, 4}

\usepackage[most]{tcolorbox}
\usepackage{csquotes}
\newtcolorbox{myquote}[1][]{%
    colback=black!5,
    colframe=black!5,
    notitle,
    sharp corners,
    borderline west={2pt}{0pt}{black!80!black},
    enhanced,
    breakable,
    top=0.5pt,
    bottom=0.5pt
}

\newcommand{\myNum}[1]{(\emph{#1})}

\begin{document}
\def\BibTeX{{\rm B\kern-.05em{\sc i\kern-.025em b}\kern-.08em
    T\kern-.1667em\lower.7ex\hbox{E}\kern-.125emX}}
\title{What Users Value and Critique: Large-Scale Analysis of User Feedback on AI-Powered Mobile Apps}

\newcommand{\stitle}[1]
{\noindent\textup{\textbf{#1}}}
\newcommand\todou[1]{\textcolor{red}{\textit{Umar: #1}}}
\newcommand\todok[1]{\textcolor{blue}{\textit{Krishna: #1}}}
\newcommand\todov[1]{\textcolor{blue}{\textit{Vinaik: #1}}}

\newenvironment{userquote}
{\begin{quote}\itshape\color{blue!60!black}}
{\end{quote}}

\author{\IEEEauthorblockN{Vinaik Chhetri\IEEEauthorrefmark{1}, Krishna Upadhyay\IEEEauthorrefmark{1}, A.B. Siddique\IEEEauthorrefmark{2}, Umar Farooq\IEEEauthorrefmark{1}}
\IEEEauthorblockA{\IEEEauthorrefmark{1}
    Louisiana State University, 
    \IEEEauthorrefmark{2} 
    University of Kentucky \\
 Email: 
vchhet2@lsu.edu, kupadh4@lsu.edu, siddique@cs.uky.edu, ufarooq@lsu.edu
}}

\maketitle
\thispagestyle{plain}
\pagestyle{plain}
\begin{abstract}

Artificial Intelligence (AI)-powered features have rapidly proliferated across mobile apps in various domains, including productivity, education, entertainment, and creativity. 
However, how users perceive, evaluate, and critique these AI features remains largely unexplored, primarily due to the overwhelming volume of user feedback.
In this work, we present the first comprehensive, large-scale study of user feedback on AI-powered mobile apps, leveraging a curated dataset of 292 AI-driven apps across 14 categories with 894K AI-specific reviews from Google Play. 
We develop and validate a multi-stage analysis pipeline that begins with a human-labeled benchmark and systematically evaluates large language models (LLMs) and prompting strategies.  
Each stage, including review classification, aspect-sentiment extraction, and clustering, is validated for accuracy and consistency.
Our pipeline enables scalable, high-precision analysis of user feedback, extracting over one million aspect–sentiment pairs clustered into 18 positive and 15 negative user topics.
Our analysis reveals that users consistently focus on a narrow set of themes: positive comments emphasize productivity, reliability, and personalized assistance, while negative feedback highlights technical failures (e.g., scanning and recognition), pricing concerns, and limitations in language support.
Our pipeline surfaces both satisfaction with one feature and frustration with another within the same review. These fine-grained, co-occurring sentiments are often missed by traditional approaches that treat positive and negative feedback in isolation or rely on coarse-grained analysis. 
To this end, our approach provides a more faithful reflection of the real-world user experiences with AI-powered apps.
Category-aware analysis further uncovers both universal drivers of satisfaction and domain-specific frustrations. 
We expect our findings to advance understanding of user-centered development of AI-powered features and provide actionable guidance to software engineers and app developers who seek to align AI features of apps with user expectations. 

\end{abstract}


\maketitle
\section{Introduction}
\label{sec:intro}
Artificial Intelligence (AI)-powered features are rapidly becoming standard in mobile apps.
In 2024 alone, mobile apps with AI features surpassed 17 billion downloads and generated \$1.3 billion in revenue~\cite{state-of-mobile-ai}. 
Once confined to chatbots and assistants, AI-driven features such as visual recognition, generative art, and recommendations are now embedded across a wide range of app categories, including productivity, entertainment, education, communication, and creative tools. 
This unprecedented growth presents significant new opportunities as well as challenges for software engineers, product managers, and user experience researchers.

\begin{figure}[t!]
  \centering
  \includegraphics[width=\columnwidth]{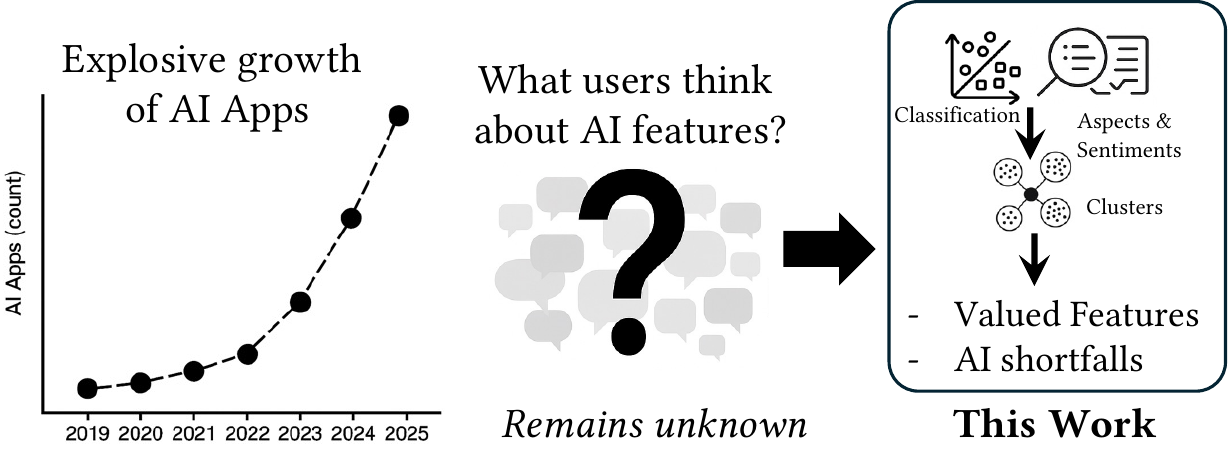}
  \caption{Despite the explosive growth of AI-powered mobile apps, user perspectives on AI-based features remain largely unknown. This work addresses this gap by automatically extracting and clustering aspects from large-scale user reviews, revealing both valued AI capabilities and shortfalls. 
  }
  \label{fig:method-overview}
  \vspace{-15pt}
\end{figure}

While technical benchmarks and performance metrics for AI models are abundant, evidence on the reception of these features ``in the wild'' is notably scarce (see Fig.~\ref{fig:method-overview}). 
User reviews, despite being a primary feedback channel for developers, remain an underutilized resource for assessing the practical impact, strengths, and limitations of AI-driven capabilities in everyday use.
Although prior research has mined user reviews to uncover usability issues, security and privacy problems, feature requests, bugs, and general user sentiment in conventional mobile apps~\cite{privacy-at-scale, pprior, tushev-domain-specific, pagano-feedback, devine-what, herman-app-store}, empirical studies focused specifically on AI-powered features remain limited.
Recent work has begun to address AI transparency and explainability, but these studies are typically narrow in scope -- often focusing on single app categories or isolated features~\cite{Fairness, li-empirical-ai}. 
Critically, there is a lack of large-scale systematic analyses that capture the breadth of AI capabilities now integrated in mobile apps.
As a result, we still know little about what users value or critique in AI-based features, which themes dominate feedback, or how user expectations are evolving as AI becomes pervasive. 
To fill this gap, we present the first comprehensive, large-scale study of user feedback on AI-driven mobile apps, with category-aware analysis spanning diverse application domains.

\stitle{Overview of this Work.}
In this work, we focus on AI-driven mobile apps, \emph{defined as apps that explicitly state in their Google Play description that they leverage AI -- such as large language models (LLMs), computer vision, or deep learning -- to power app features.}
As Fig.~\ref{fig:method-overview} shows, our pipeline begins with curation and manual vetting of 292 AI-driven apps from the Google Play, spanning 14 app categories and yielding a dataset of over $2.2M$ user reviews after data cleaning. 
To distinguish truly AI-related reviews from generic app feedback, we develop a human-labeled benchmark, applying annotation guidelines and inter-rater agreement checks.
Leveraging this benchmark, we systematically evaluate a broad set of open-source LLMs -- including Llama-3~\cite{llama3herdmodels}, Mistral~\cite{mistralai_mistral7b_instruct_v0.1}, Mixtral~\cite{mixtral-8x7B-Instruct-v0.1}, and DeepSeek~\cite{deepseekai2025deepseekr1incentivizingreasoningcapability} -- in combination with a diverse range of prompting strategies, such as rule-based, chain-of-thought, in-context examples, and app-specific feature grounding. 
Our experiments show that Llama3 70B, when prompted with in-context examples, app-specific AI features, and chain-of-thought reasoning, achieves 94.4\% F1 score in AI-related review classification. 
Llama3 70B's strong performance allows us to extend the labeling process to the entire dataset, resulting in the identification of over 890K reviews (out of 2.2M) that explicitly reference AI features.

Once AI-related reviews are identified, we apply our best-performing LLM configuration to extract aspect–sentiment pairs, where each aspect refers to a specific AI-powered feature, capability, or user-facing function mentioned in the review, and the sentiment indicates the user's subjective judgment (i.e., positive, negative, or neutral) toward that aspect.
This process achieves over 92\% accuracy in human validation and generates more than 1.1 million such pairs.
To uncover latent themes in this vast corpus, we represent each aspect using sentence embeddings~\cite{mpnet, all-mpnet-base-v2} and apply k-means clustering. 
This analysis yields 18 positive and 15 negative user topics, each supported by quantitative prevalence statistics and illustrative review excerpts.

\begin{figure}[t!]
  \centering
  \includegraphics[width=\columnwidth]{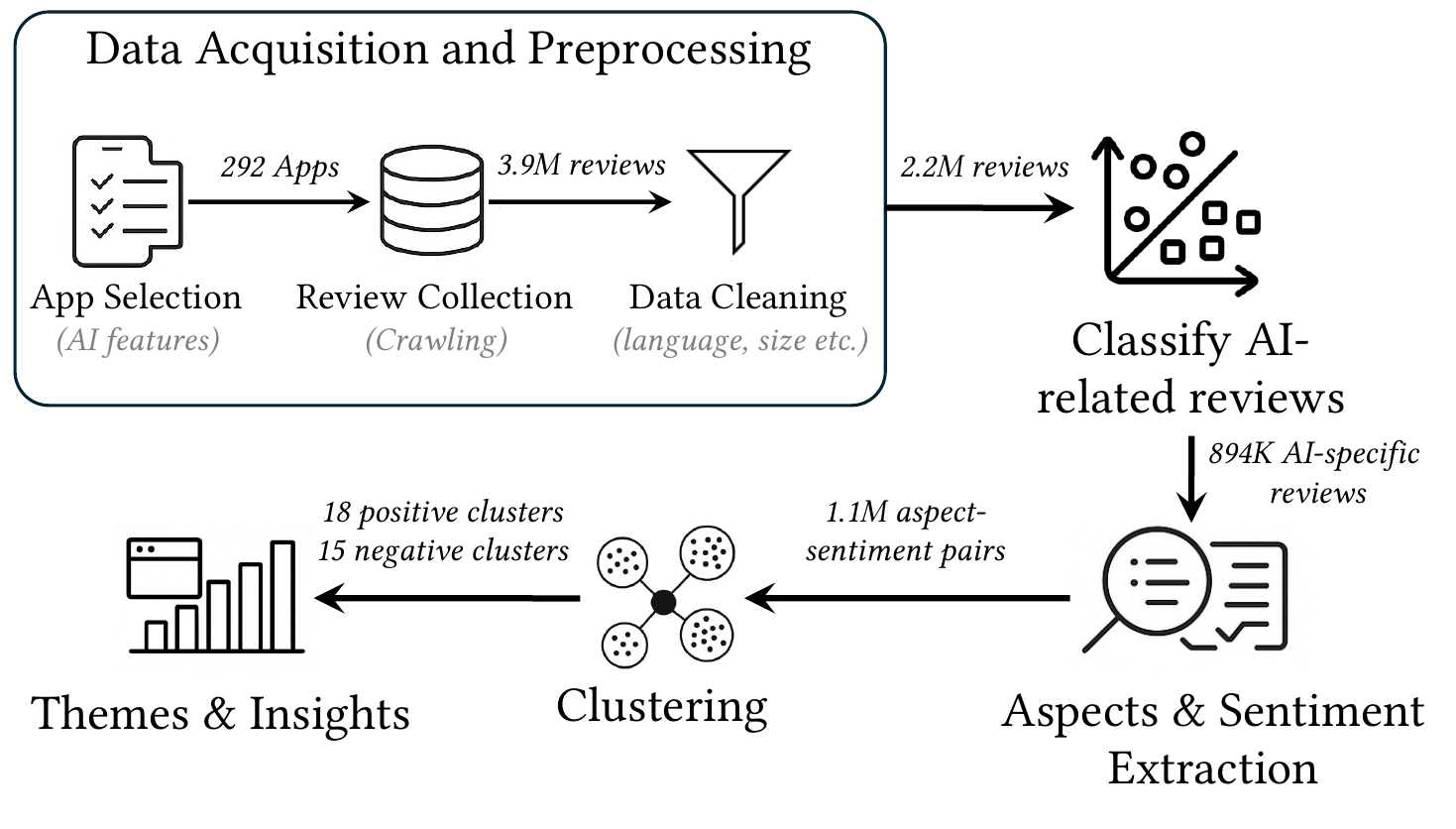}
  \caption{Overview of our approach. The pipeline begins with Data Acquisition \& Preprocessing (i.e., app selection, review collection, data cleaning, and feature extraction), followed by LLM-based classification, aspect and sentiment extraction, clustering, and thematic analysis to generate actionable insights.}
  \label{fig:method-overview}
  \vspace{-15pt}
\end{figure}

Our findings offer a detailed characterization of the current state of user experience with AI features in mobile apps.
Positive feedback consistently highlights productivity enhancement, intelligent assistance, and personalization as the features that deliver the greatest value to users.

In contrast, negative feedback focuses on technical failures in scanning, recognition, and math, as well as concerns about pricing and limitations in language support.
Importantly, our analysis captures both satisfaction and frustration within the same review, which traditional methods often overlook, providing a more faithful reflection of real-world user experiences with AI-powered apps.
Our category-aware analysis further uncovers both universal trends (such as widespread demand for robust and reliable AI features) and domain-specific expectations, including precise math solutions for education, creative flexibility for art and design, and multilingual fluency for communication tools. 
Drawing on these insights, we recommend that developers prioritize resolving common pain points associated with valued features by improving the reliability and usability of core AI functions, while also customizing AI capabilities to the distinct requirements of each app category.

\stitle{Key Contributions.} 
This work makes the following contributions:

\begin{itemize}
\item We construct and release the first benchmark dataset of 292 AI-driven apps across 14 categories, featuring 894K user reviews that explicitly reference AI-based features.

\item We design and implement an automated pipeline that uses state-of-the-art LLMs to classify reviews, extract fine-grained aspect–sentiment pairs, and organize user feedback at scale.

\item Our analysis identifies 18 positive and 15 negative user topics by clustering aspect–sentiment pairs using semantic embeddings, revealing key areas of user satisfaction and pain points.

\item Our dataset, LLM-based pipeline code, and clustering artifacts are made publicly available for future research.
\end{itemize}

\section{Background} \label{sec:background}
\stitle{AI in Mobile Apps.}
The integration of AI-based features into mobile apps has advanced rapidly, enabling a wide range of intelligent functionalities. 
Modern apps now deliver personalized content, context-aware recommendations, natural language understanding, healthcare analytics, real-time translation, accessibility enhancements, and creative generative capabilities such as image synthesis and style transfer~\cite{empirical-ai}. 
As AI becomes more deeply embedded into mobile apps, understanding the resulting user experience has become increasingly important.

Unlike traditionally programmed features, AI-driven functionalities are inherently adaptive and data-dependent, often resulting in behavior that appears opaque or unpredictable to end users. This opacity underscores the importance of user feedback, as reviews reflect real-world perceptions and can surface concerns such as perceived fairness, privacy risks, and security vulnerabilities.
Recent empirical studies have begun to address these concerns, investigating fairness perceptions in AI-enabled apps through user feedback~\cite{Fairness}, analyzing trends in app characteristics and sentiment~\cite{empirical-ai}, and identifying security vulnerabilities in LLM-based mobile apps~\cite{llmsecurity}.

AI-based features are no longer an add-on in mobile apps; they have become central to core functionalities. 
As frameworks like TensorFlow Lite and Core ML simplify the integration of advanced AI features, it becomes essential to systematically analyze how users perceive, interact with, and evaluate these capabilities. 
A clear understanding of user feedback is critical for informing the design, debugging, and refinement of AI-driven functionality in real-world mobile apps.

\stitle{User Reviews as a Source of Feedback.}
User reviews in mobile app stores offer developers a valuable lens into real-world user experiences, enabling the identification of well-received features, pain points, and opportunities for improvement~\cite{Bug,noisy-rev-sum}. 
Leveraging this feedback can directly inform app evolution and enhance both user satisfaction and software quality~\cite{core,noisy-rev-sum}. 
However, effectively utilizing user reviews poses several challenges. 
The sheer volume and scale of incoming feedback can overwhelm manual inspection~\cite{aarsynth,core,noisy-rev-sum}, while noise, redundancy, and the prevalence of off-topic (e.g., discussion of non AI-powered features) or uninformative content further complicate analysis~\cite{arminer,noisy-rev-sum,Automatic-Summarization-of-Helpful-App-Reviews}. 
Most app reviews are short and informal, and studies estimate that only about 35\% of reviews provide informative or actionable content~\cite{arminer}. Consequently, extracting concise, sentiment-aware, and relevant feedback at scale remains a persistent challenge for both researchers and practitioners~\cite{Automatic-Summarization-of-Helpful-App-Reviews}.

AI-specific user feedback poses many of the same challenges, often compounded by the implicit and rapidly evolving nature of AI-driven features. In this work, we propose a scalable pipeline for surfacing and synthesizing insights about AI-related features from large-scale, noisy user review data.

\section{Study Design}
\label{sec:study-design}

\begin{table}[t]
\centering
\caption{Overview of the AI-related mobile app review data.}
\label{tab:dataset-stats}
\begin{tabular}{lr}
\toprule

Number of Apps & $292$ \\
Number of App Categories & $14$\\
Number of reviews (total) & $3.9M$ \\
Number of review (post-cleaning) & $2.2M$ \\
Number of AI-related reviews & $894K$\\
\bottomrule
\end{tabular}
\vspace{-15pt}
\end{table}

\subsection{Study Objective and Research Questions}
Our approach is inspired by the Goal Question Metric (GQM) framework~\cite{gqm} and the multi-stage process for research question derivation adopted in~\cite{nahar2024productbeyond}. 
We prioritized questions that are both of high practical relevance and amenable to rigorous, data-driven analysis.
We focus on the following research questions:

\begin{myquote}
\textbf{RQ1:} \textit{What are the most common user concerns and user-reported benefits associated with AI-powered features in mobile apps?}
\end{myquote}
\noindent\textit{Motivation:} To empirically characterize the range and prevalence of positive and negative user feedback, moving beyond anecdotal evidence and single-case studies.

\begin{myquote}
\textbf{RQ2:} \emph{How do user-valued features and reported problems co-occur within the same review?}
\end{myquote}
\noindent\emph{Motivation:} Many user experiences are mixed rather than purely positive or negative; analyzing praise-problem co-occurrence reveals critical trade-offs and highlights where improvements are most needed.

\begin{myquote}
\textbf{RQ3:} \emph{How do AI-based features vary across different app categories?}
\end{myquote}
\noindent\emph{Motivation:} Different domains may face unique adoption challenges and opportunities; domain-specific analysis can inform targeted improvements and transferability.

\begin{table}[t]
\centering
\caption{Reviews distribution across app categories.}
\label{tab:review-distribution}
\begin{tabular}{lrr}
\toprule
\textbf{App Category} & \textbf{\# of Reviews} & \textbf{Distribution(\%)} \\
\midrule
Productivity & 263,125 & 29.43 \\
Education & 154,350 & 17.26 \\
Tools & 151,343 & 16.93 \\
Entertainment & 110,369 & 12.34 \\
Photography & 96,088 & 10.75 \\
Art \& Design & 40,458 & 4.53 \\
Personalization & 36,669 & 4.10 \\
Communication & 16,324 & 1.83 \\
Video Players \& Editors & 9,901 & 1.11 \\
Music \& Audio & 4,506 & 0.50 \\
Books \& Reference & 3,939 & 0.44 \\
Health \& Fitness & 3,590 & 0.40 \\
Lifestyle & 2,178 & 0.24 \\
Travel \& Local & 1,223 & 0.14\\
\bottomrule
\end{tabular}
\vspace{-15pt}
\end{table}

\subsection{Selection of Apps and Reviews}
We define an app as \emph{AI-driven} if the ``About this app'' section explicitly advertises at least one feature using AI, machine learning, natural language processing, computer vision, or comparable AI technologies (e.g., AI-powered document scanner, ML-based photo enhancement, or intelligent text summarization).

Building on this definition, we assembled a comprehensive set of candidate AI-driven mobile apps (e.g., ChatGPT~\cite{chatgpt}, Gemini~\cite{gemini}) from the Google Play Store~\cite{googleplay} through a structured, multi-phase search. 
We began with a seed set of prominent apps recognized for their AI capabilities and expanded the pool by recursively traversing ``similar apps'' recommendations across diverse categories such as Productivity, Education, Health, Finance, and Communication.

We initially identified $310$ candidate mobile apps mentioning AI features. 
Following manual review of each app's ``About this app'' section to verify the presence of AI-driven features, the set was refined to $292$ unique AI apps spanning $14$ categories. 
From these apps, we collected approximately $3.9M$ user reviews. In line with established practices~\cite{rrgen,aarsynth}, we conducted several data cleaning steps:
\myNum{i}~exclusion of reviews with fewer than five words \cite{Ebrahimi}; and
\myNum{ii}~filtering for English-language content using automated language detection.
The resulting dataset comprises $2.2M$ reviews, each mapped to an app meeting our explicit criteria for AI-driven functionality.
Table~\ref{tab:dataset-stats} presents statistics of our curated data and Table~\ref{tab:review-distribution} shows the distribution of AI-related reviews across categories of the apps.

\subsection{Human-Labeled Benchmark for Review Classification}
Since mobile app reviews frequently include feedback on a wide range of general features, it is necessary to first identify and extract only those reviews that specifically discuss AI-powered functionalities to meaningfully analyze user perspectives on AI-related aspects. 
To establish a reliable evaluation benchmark for this filtering step, we randomly sampled over 300 user reviews from the curated data to construct a high-quality evaluation benchmark for AI relevance classification. 
Two authors independently annotated each review as either \emph{AI-related} or \emph{non-AI-related}, based on whether the review explicitly referenced an app's AI-powered functionality. 
Agreement between annotators was measured using Cohen's kappa~\cite{cohen1960coefficient}, 
which indicated a substantial agreement ($\kappa=0.7139$). 
For the benchmark, we retained only those reviews where both annotators independently assigned the same label (i.e., 270 reviews), resulting in a set with unambiguous ground truth for subsequent model evaluation.

\section{Methodology}
\label{sec:method}
In this section, we describe how we leverage state-of-the-art 
LLMs to automatically classify, extract, and cluster AI-specific user feedback for mobile apps.

\subsection{Review Labeling using LLM}
To automatically classify whether a given user review discusses any AI-related feature or not, we employ open-source LLMs and reproducible prompt design.
We systematically experimented with a range of LLMs and prompting strategies to identify the most effective configuration for automated labeling.

\subsubsection{Models}
\label{subsubsec:models}
We selected a suite of state-of-the-art open-source LLMs representing architectural diversity (including dense and Mixture-of-Experts variants) developed by multiple research organizations. The evaluation set included models from Mistral, Meta, and DeepSeek, with parameter counts spanning from $7B$ to $70B$. 
This selection enables rigorous comparison across model families, sizes, and distillation approaches.

\begin{table*}[t!]
\centering
\footnotesize
\caption{Model performance (F-1 score) comparison across different instruction-tuned models.}
\resizebox{\textwidth}{!}{%
\begin{tabular}{l|c|cc|cc|cccc}
\toprule
\multirow{2}{*}{Prompt-type} & \multicolumn{1}{c|}{Mistral} & \multicolumn{2}{c|}{Mixtral} & \multicolumn{2}{c|}{Llama} & \multicolumn{4}{c}{DeepSeek-R1-Distill}  \\
\cmidrule(l{0pt}r{0pt}){2-2} \cmidrule(l{0pt}r{0pt}){3-4} \cmidrule(l{0pt}r{0pt}){5-6} \cmidrule(l{0pt}r{0pt}){7-10}
 & 7B & 8x7B & 8x22B  & 8B & 70B & Qwen-7B & Qwen-32B    & Llama-8B & Llama-70B \\
\midrule
Zero-shot (naive) & 0.5556 & 0.7867 & 0.9013 & 0.2535   & 0.8996  & 0.8400 & 0.9289 & 0.8724  & 0.9243 \\
Zero-shot (naive) + CoT & 0.5856 & 0.7512 & 0.9205 & 0.6944 & 0.9129 & 0.8502 & 0.9300 & 0.8630  & 0.9383   \\
\midrule
Zero-shot (Rule based) & 0.6011 & 0.6304 & 0.7600 & 0.7892  & 0.7864 & 0.7610 & 0.7662 & 0.8019  & 0.8545   \\
Zero-shot (Rule based)  + CoT & 0.6383 & 0.6943 & 0.7685 & 0.7685   & 0.8584 & 0.7228 & 0.7805 & 0.7692  & 0.8597   \\
\midrule
In-Context Examples & 0.6221 & 0.8720 & 0.8906 & 0.8077   & 0.9147  & 0.8133 & 0.9160 & 0.7981 & 0.9163  \\
+ CoT (1 example) & 0.7200 & 0.8293 & 0.8108 & 0.8205  & 0.9042 & 0.7792 & 0.9042 & 0.8333   & 0.9249   \\
+ CoT (2 examples) & 0.6951 & 0.8309 & 0.8655 & 0.8178  & 0.8839 & 0.7838 & 0.9119 & 0.8745 & 0.9125    \\
+ CoT (3 examples) & 0.6610 & 0.8514 & 0.8163 & 0.7785   & 0.9030 & 0.7958 & 0.8889 & 0.8692  & 0.8981   \\
\midrule
In-Context Examples + App specific features & 0.6253 & 0.8699 & 0.9057  & 0.8100  & 0.9333 & 0.7822 & 0.9052 & 0.7788  & 0.9417   \\
+ CoT (1 example) & 0.7103 & 0.8340 & 0.7907 & 0.8172  & \textbf{0.9444} & 0.6188 & 0.9339 & 0.7678   & 0.9306   \\
+ CoT (2 examples) & 0.6630 & 0.8417 & 0.8247 & 0.8227 & 0.9057 & 0.7966 & 0.9048 & 0.8538  & 0.9183    \\
+ CoT (3 examples) & 0.6667 & 0.8538 & 0.7792 & 0.8108  & 0.9008 & 0.8390 & 0.9286 & 0.8538    & 0.9255  \\
\bottomrule
\end{tabular}
}
\label{tab:model_comparison}
\vspace{-15pt}
\end{table*}

\subsubsection{Prompting Strategies}
\label{subsubsec:prompts}
Prompt design plays a critical role in eliciting accurate and contextually grounded responses from LLMs. 
To rigorously evaluate the sensitivity of LLM-based review classification to prompt formulation, we designed and compared twelve distinct prompting strategies, each targeting a specific axis of model reasoning, instruction-following, and context utilization.

\stitle{Zero-shot naive prompting.}
This approach presents the model with the review text and app name only, asking whether the review is AI-related or not. It involves no explicit instructions or demonstrations.

\stitle{Zero-shot rule-based prompting.}
We augment the naive prompt with explicit decision rules that define what constitutes an AI-related review. The prompt provides a list of criteria (i.e., heuristics) for the model to consider (e.g., ``If the review mentions machine learning, deep learning, or smart features, consider it AI-related''). 
This approach tests whether explicit operationalization of the task improves accuracy without in-context examples.

\stitle{Chain-of-Thought (CoT) prompting.}
We adapt both naive and rule-based prompts by appending ``Let’s think step by step.'' This encourages the model to articulate its reasoning process before making a final classification. The CoT strategy is motivated by evidence that LLMs, when prompted to perform intermediate reasoning steps, often achieve greater factual accuracy and robustness~\cite{kojima2023largelanguagemodelszeroshot}.

\stitle{In-Context Examples.}
This prompting strategy provides several labeled examples of reviews and their correct classifications, allowing the model to learn from these demonstrations. 
The selected examples span a range of review phrasings and levels of explicitness, including both prototypical and borderline cases. 
This strategy leverages LLMs' in-context learning abilities, and has been shown to improve classification and generalization over zero-shot prompting~\cite{few-shot}.

\stitle{In-Context Examples with CoT.}
We further augment the in-context example prompts by including reasoning chains for each demonstration review (e.g., ``This review is AI-related because it refers to the app's intelligent recommendation system''). 
The model is encouraged to generate both a classification and a supporting rationale.

\stitle{In-Context Examples with App-Specific AI Features.}
LLMs often hallucinate the presence or absence of AI functionality based solely on the app name, especially when the name contains the term with some relevance to ``AI.''
This behavior was especially evident when Chain-of-Thought (CoT) reasoning was enabled—the intermediate reasoning revealed that the model was making incorrect assumptions about the app’s capabilities based solely on its name and domain. For example, if the app was related to music or photography, the model often inferred the presence of sophisticated AI features typical of those domains (e.g., music generation or image enhancement), even when such features were not mentioned or did not exist.
To mitigate this, we explicitly provide the structured AI feature list for the app as additional context within the prompt. The model is instructed to reference these documented features when determining if a review discusses an AI-driven aspect. This approach grounds classification decisions in factual app capabilities and reduces reliance on heuristics. 
This approach is inspired by prior work~\cite{Ebrahimi, aarsynth}, which demonstrated the effectiveness of incorporating domain- and app-specific context to enhance the accuracy of review summarization and automated response generation.

\stitle{In-Context Examples + App-Specific AI Features + CoT.}
The most context-rich prompting strategy combines multiple labeled demonstrations, explicit app feature lists, and stepwise reasoning chains. This formulation is designed to minimize ambiguity, provide maximum factual grounding, and elicit the most stable and accurate model behavior. 
It enables the LLM to reason not just from textual clues in the review but also from structured, app-level evidence.
This strategy was motivated by prior work on LLM prompting~\cite{kojima2023largelanguagemodelszeroshot, few-shot} and was designed to address challenges such as implicit references, model hallucination, and ambiguity in user language.

\stitle{Comparison of Models and Prompting Strategies.}
Across all prompting strategies, we evaluated model performance using the human-labeled benchmark introduced earlier. As shown in Table~\ref{tab:model_comparison}, Llama-70B achieved the highest F-1 score when prompted with a single in-context example, explicit app-specific features, and CoT reasoning. In addition to its best performance, Llama-70B demonstrated consistent accuracy across multiple configurations, indicating strong generalization and robustness. 
These qualities make it particularly well-suited for large-scale review labeling, where model consistency and reliability are critical.
Interestingly, adding more in-context examples (two or three) for Llama-70B degraded performance compared to a single example. 
This decline can be attributed to significantly lengthy prompt, which may dilute the relevance of each example and hinder the model's ability to focus on the most salient patterns. 
This also suggests that for large instruction-tuned models (e.g., Llama-70B, Qwen-32B), a well-crafted example combined with app-specific features and CoT reasoning is more effective than multi-shot prompting.

Guided by these results, we leveraged Llama-70B with the best-performing prompting strategy to classify the full corpus of $2.2M$ reviews, resulting in the identification of $894K$ (41\%) reviews as AI-related for the subsequent aspect and sentiment extraction step.

\begin{figure}[t!]
\centering
\footnotesize
\begin{tcolorbox}[colframe=black!65, colback=blue!2, boxrule=0.6pt, arc=3pt, left=3pt, right=3pt, top=3pt, bottom=3pt]
\small
\textbf{Aspects and Sentiments Extraction Task Steps:}
\begin{enumerate}[leftmargin=1.2em]
    \item Extract AI-related content from the review, determine if the extracted content discusses issues or positive attributes of the app's AI features: 
    {If not, return:} 
    \texttt{\{"ai\_related": "No", "summaries": \{\}\}}
    \item Generate a dictionary of short summaries (max 3 words each) for the app's AI features, with sentiment (\texttt{"positive"}, \texttt{"negative"}, or \texttt{"neutral"}).
    \item Return JSON: 
    {If \texttt{"ai\_related"} is \texttt{"Yes"}, the \texttt{"summaries"} dictionary contains short summaries (keys) and corresponding sentiment (values):} \\
    \texttt{\{"ai\_related": "Yes", "summaries": \{"<short\_summary>": "<sentiment>"\}\}} \\
    {If not, return:} 
    \texttt{\{"ai\_related": "No", "summaries": \{\}\}}
\end{enumerate}

\textbf{Instructions:}
\begin{itemize}[leftmargin=1.2em]
    \item A review is \texttt{"AI-related"} only if it discusses the app's AI features. \textcolor{blue}{Simply mentioning AI keywords is not enough.}
    \item {Summaries must be concise (max 3 words), strictly AI-specific, and ignore non-AI topics.}
    \item Sentiment: \texttt{"positive"}, \texttt{"negative"}, or \texttt{"neutral"}.
    \item {Output must be valid JSON with empty fields if applicable.}
\end{itemize}

\textbf{Example:}\\
{\textbf{App name:}} Brainly: AI Homework Helper\\
\textbf{Features:} Quick Math Solver, Step-by-Step Solutions ... \\
{\textbf{Review:}} \textit{Ok, 1st Q: Is the earth flat? ...}

\textbf{Step-by-step:}
\begin{enumerate}[leftmargin=1.7em, itemsep=0pt]
    \item Extract AI-related content: \textit{Mentions Brainly's answer system (AI-related).}
    \item Is it related to the app's AI features? \textit{Yes; limitation in Quick Math Solver.}
    \item Short summary: \texttt{"message length limitations": "negative"}
    \item JSON output:
\end{enumerate}
\texttt{\{"ai\_related": "Yes", "summaries": \{"message length limitations": "negative"\}\}}

\vspace{0.3em}
\textbf{Input:}\\
{The app offers the following AI features: \textcolor{blue}{\{\texttt{feature}\}}. name of the mobile app is \textcolor{blue}{\{\texttt{app\_name}\}} and classify the following mobile app review accordingly: \textcolor{blue}{\{\texttt{review\_text}\}}}
\end{tcolorbox}
\caption{LLM prompt for extracting aspects and sentiments.}
\label{fig:aspect_prompt}
\vspace{-20pt}
\end{figure}

\subsection{Aspect and Sentiment Extraction}
\label{subsec:aspect-extraction}

In this step, we extract fine-grained aspect–sentiment pairs from AI-related reviews, which enables systematic analysis of user feedback beyond coarse sentiment labels. 
Unlike prior work~\cite{privacy-at-scale}, which represents each review as a single embedding, our approach decomposes each review into multiple distinct aspect–sentiment units. 
This allows a single review to contribute to multiple themes, leading to improved clustering quality, particularly when reviews contain mixed opinions.

We used Llama-70B, the best-performing model from our earlier evaluation, along with a specialized prompt designed to elicit both aspect categories and associated sentiment labels from each review (see Figure~\ref{fig:aspect_prompt}). 
We evaluated performance on the 128 AI-related reviews from our human-labeled benchmark. 
The model correctly identified 122 of these as AI-related and extracted a total of 213 aspect–sentiment pairs, as many reviews referenced more than one AI-related feature or concern.

\setlength{\tabcolsep}{3pt}
\renewcommand{\arraystretch}{1}
\begin{table*}[]
\footnotesize
\centering
\caption{Percentage improvement in Silhouette Score for positive and negative clusters as the number of clusters 
\(k\) increases.}
\vspace{-5pt}
\label{tab:kmeans-eval}
\begin{tabular}{lrrrrrrrrrrrrrrrrrr}
\hline
$\mathbf{k}$ & 3 & 4 & 5 & 6 & 7 & 8 & 9 & 10 & 11 & 12 & 13 & 14 & 15 & 16 & 17 & 18 & 19 & 20 \\ \hline 
\textbf{Negative} & 0.000 & 0.000 & 2.955 & 12.575 & 0.000 & 4.109 & 5.271 & 4.297 & 2.928 & 2.980 & 2.942 & 1.669 & 2.200 & 0.000 & 0.000 & 0.000 & 0.000 & 0.000 \\ \hline 
\textbf{Positive} & 69.120 & 28.149 & 12.485 & 2.310 & 3.515 & 4.096 & 5.904 & 5.278 & 0.000 & 4.039 & 0.000 & 3.743 & 1.225 & 2.021 & 0.000 & 1.002 & 0.000 & 0.000 \\ \hline
\end{tabular}
\vspace{-15pt}
\end{table*}

To assess extraction quality, two human annotators independently reviewed all 213 aspect–sentiment pairs and agreed that 196 were valid, indicating that 92\% of the model’s extractions were accurate.
This high level of accuracy demonstrates that our pipeline produces reliable results for downstream analysis.

After validating model performance, we applied the same approach to the entire set of $894K$ AI-related reviews. 
The aspect–sentiment extraction step yielded over $1.1M$ million pairs from $894K$ AI-related reviews. 
This step converts unstructured user reviews into structured aspect–sentiment pairs, enabling a more granular view of user feedback. 
This representation captures fine-grained opinions, enabling each review to reflect multiple features or concerns, and even both satisfaction and frustration in the same instance.
Critically, it preserves the coexistence of praise and criticism within the same review, offering a richer and more realistic understanding of user experience. 
Clustering is therefore essential to group semantically similar expressions, reduce noise, and distill coherent themes that summarize common patterns of satisfaction and frustration with AI-powered app features.

\subsection{Clustering Review Aspects} 
\label{subsec:clustering}

In this step, we cluster the extracted aspect–sentiment pairs to identify high-level themes that summarize common patterns in user feedback.

\subsubsection{Sentiment-Based Partitioning}
To enhance semantic clarity and interpretability of the clusters, we partitioned the aspect–sentiment pairs by sentiment polarity (positive vs. negative) before clustering. 
Without this separation, opposing views about the same feature, for instance, ``fast and seamless voice recognition'' and ``voice commands rarely work'', could be grouped into a single cluster due to topical similarity.  
Additionally, many aspects, such as ``translation'' occur in both positive and negative contexts.
This could conflate user satisfaction or frustration, making it harder to extract clear, actionable insights. 
By isolating positive and negative aspect instances, we preserved sentiment-specific nuance, e.g., positive feedback on ``translation'' remains distinct from criticism of the same feature. 
This partitioning enables more accurate downstream analysis of user perception. 

\subsubsection{\(k\)-Means Clustering}
After partitioning, we reduced the $1.1M$ aspect–sentiment pairs to $168K$ (168,283) unique aspects, consisting of $70K$ (70,119) positive and $98K$ (98,164) negative instances.
We embedded each aspect using the \texttt{sentence-transformers/all-mpnet-base-v2}~\cite{mpnet, all-mpnet-base-v2} model and applied \(k\)-means clustering separately for positive and negative sets. 
To determine the optimal number of clusters, we evaluated the quality of clusters using the average Silhouette Score~\cite{silhoutte} as the evaluation metric for different values of \(k\), incrementally increasing \(k\) until no improvement was observed across several consecutive values.
We used the Silhouette Score as it balances intra-cluster cohesion and inter-cluster separation, making it a well-suited metric for evaluating the compactness and distinctiveness of clusters in high-dimensional semantic spaces.

\subsubsection{Cluster Evaluation}
Clustering performance for 
\(k= 3 \cdots 20 \) is summarized in Table~\ref{tab:kmeans-eval}, which reports the percentage improvement in clustering quality relative to the best score observed up to each \(k\).
Specifically, at cluster $k$:
\[
S_{\max}(k) = \max_{i=1,2,\ldots,k} S(i),
\]
where $S(i)$ is the silhouette score at cluster $i$.

\[
\text{Improvement}_k = \frac{S_{\max}(k) - S_{\max}(k-1)}{|S_{\max}(k-1)|} \times 100
\]

For negative aspects, we observed a pronounced increase in Silhouette Score for \(k=6\), with no improvements beyond \(k=15\). 
We therefore selected \(15\) as the optimal number of clusters for negative sentiment. 
For positive aspects \(k=18\) was chosen, corresponding to the last observed improvement in Silhouette Score. 
This procedure resulted in 15 negative and 18 positive clusters, each capturing a coherent, sentiment-specific theme in the user feedback.

\section{Analysis and Insights} 
\label{sec:themes}
Table~\ref{tab:prominent-topics} presents the distribution of positive and negative clusters identified from user reviews.
We identified 18 positive and 15 negative clusters, representing 70,119 unique positive aspects 
and 98,164 unique negative aspects.
Notably, 41,499 reviews included both positive and negative aspects, highlighting the multifaceted nature of user feedback on AI-powered mobile apps.
Note that a single review can discuss multiple aspects and consequently contribute to several clusters.
The following sub-sections provide a detailed analysis of the most prominent positive and negative topics, exploring key patterns in user sentiment and priorities across the AI-based mobile apps.

\begin{table*}[t]
\centering
\footnotesize
\caption{Distribution and Descriptions of Positive and Negative Topics in User Reviews.}
\label{tab:prominent-topics}
\begin{tabular}{@{}p{6.6cm}rr|p{6.6cm}rr@{}}
\toprule
\textbf{Positive Topic (Description)} & \textbf{Aspects} & \textbf{Reviews} & \textbf{Negative Topic (Description)} & \textbf{Aspects} & \textbf{Reviews} \\
\midrule
\textbf{AI Assistant:} Versatile AI companion that provides personalized help, problem-solving, and interactive support across multiple tasks and domains. & 6,949 & 48,999 & \textbf{Scanner Issues:} Scanner Issues. & 13,208 & 62,598 \\

\textbf{ScanToSolve:} Reliable Scan-to-Solution Support for Fast and Clear Problem Solving. & 6,260 & 30,680
& \textbf{Access \& Pricing:} Cost, Access Limitations and Subscription and Paywall Frustrations. & 12,526 & 31,927\\

\textbf{Smart \& Personalized:} Personalized AI Tools for Easy and Smart Assistance. & 5,988 & 49,682
& \textbf{AI understanding:} AI explanation and understanding limitations. & 9,932 & 29,450\\

\textbf{Visual-Solver:} Visual Learning and Problem Solving Through Smart Photo and Video Recognition. & 5,130 & 25,347
& \textbf{AI Misbehavior:} Ineffective and misuse of AI. & 7,610 & 25,974\\

\textbf{Helpful Answer:} Accurate, Clear, Reliable and Helpful Answers \& Explanations. & 4,824 & 46,886
&\textbf{Voice Limitations:} Voice Interaction and Audio Experience Limitations. & 6,945 & 38,999\\

\textbf{AI Writing:} AI-Powered Writing, Editing, and Grammar Support for Improved Text Creation. & 4,634 & 28,429
& \textbf{OCR Issues:} Text \& Symbol Recognition Issues. & 6,540 & 25,532\\

\textbf{Creative Roleplay:} Interactive Character Creation and Roleplay for Fun, Storytelling, and Expression. & 4,402 & 27,369
&\textbf{Vision Issues:} Visual Feature Limitations and Recognition Issues. & 6,136 & 13,591\\

\textbf{Productivity:} Smart, Fast, and Accurate Solutions for Learning, Creativity, and Digital Productivity. & 4,171 & 36,851
&\textbf{Math Issues:} Inaccurate Math Solutions. & 6,080 & 36,353\\

\textbf{Helpful Guidance:} Academic Guidance, Helpful Support \& Relief. & 4,124 & 36,145
& \textbf{Chat Issues:} Chat bot bugs (malfunction) / Chatbot Performance \& Interaction Issues. & 5,409 & 22,676\\

\textbf{Immersive Chat:} Engaging and Realistic AI Chat Experience. & 3,985 & 17,976
& \textbf{Image Quality Concerns:} Photo/Image quality or Camera and Image Quality Issues. & 5,220 & 25,823\\

\textbf{Enhanced Learning:} Smarter, more effective learning using AI. & 3,628 & 40,587
& \textbf{Usage Limits:} Content, Capability, Data, and Usage Limitations. & 5,203 & 26,359\\

\textbf{AI Impressiveness:} User amazement towards AI features. & 3,428 & 23,635
&\textbf{Language Limitations:} Language Support \& Localization Issues. & 5,123 & 37,549\\

\textbf{AI Voice:} Enhanced \& Personalized Voice AI Experience. & 3,094 & 14,143
& \textbf{Slow Performance:} Slow performance. & 3,234 & 23,162\\

\textbf{Artful AI:} Creative \& Fun AI-Powered Visual Design. & 2,735 & 11,393
& \textbf{Answer Quality Issues:} Answer Quality Concerns or Issues with Answer Accuracy and Completeness. & 2,930 & 32,040\\

\textbf{Language Support:} Language Versatility in AI Systems. & 2,432 & 34,715
& \textbf{Censorship:} Content Filtering and Censorship Limitations. & 2,068 & 19,582\\

\textbf{Mathsolver:} Useful math solver. & 1,912 & 19,426 & \\
\textbf{Search:} Efficient Search Experience. & 1,566 & 10,684 & \\

\textbf{Filters:} Content Moderation \& Filtering, Aesthetic \& Fun Filter Experience. & 857 & 6,289 & \\
\bottomrule \\
\end{tabular}
\vspace{-20pt}
\end{table*}

\subsection{RQ1: Prevalence \& Distribution}
To investigate RQ1, we performed an analysis of 18 positive and 15 negative AI-related topics produced during clustering (Table~\ref{tab:prominent-topics}).
This topical organization of clusters provides a detailed account of the specific features users engage with, what they appreciate, and what frustrates them in AI-powered mobile apps.

\stitle{Positive Topics.}
The distribution of positive topics is notably concentrated in a few dominant themes. A small number of core themes, including AI Assistant, ScanToSolve, Smart \& Personalized, and Visual-Solver, emerge as the dominant points of user attention and collectively account for the majority of positive mentions in the data. 
These topics focus on delivering intelligent, context-aware assistance, rapid problem-solving, and effective task completion, as reflected in reviews such as:
\begin{userquote}
``\(\cdots\) You did great work, even very rare topics or scriptures could be searched. \(\cdots\)''
\end{userquote}

Secondary positive topics include advanced content generation (i.e., AI Writing cluster), creative or expressive features (Creative Roleplay, Artful AI).
While these are appreciated by many, their frequencies are noticeably lower, indicating that the perceived core value of AI in apps is closely tied to productivity and tangible assistance, rather than to entertainment or novelty.

\stitle{Negative Topics.}
On the negative side, user discourse is similarly focused: the largest topics (Scanner Issues, Access \& Pricing, AI understanding) capture the pain points related to technical robustness, pricing, and reliability. For example:
\begin{userquote}
    ``\(\cdots\) When u scan the, simple interest type questions he always says `Make sure you cropped only one equation. \(\cdots\)''
\end{userquote}
\begin{userquote}
    ``\(\cdots\) cool free trial but even if you pay there is 50 image per day limit (which is also pretty broken its been 24h and still limit). \(\cdots\)''
\end{userquote}

Other high-frequency topics center around problems in input recognition (OCR, Vision), calculation (Math), and communication (Voice, Chat), again showing that users' key concerns are practical and functional. 
Platform constraints (Censorship, Usage Limits), while present, are less dominant in the topic landscape.

\stitle{Breadth versus Concentration.}
A key empirical insight is the high topical concentration at both ends of the spectrum. 
The top five positive and top five negative topics together account for the majority of all user feedback related to AI features, suggesting that while the review corpus covers a diverse set of themes, users overwhelmingly focus their attention on a handful of core capabilities and shortcomings. 
This is consistent with recent large-scale studies on mobile app review mining~\cite{pagano-feedback, devine-what, herman-app-store}, where a small number of features consistently drive the most discussion and developer attention.

\stitle{Dual-Role Topics Reveal User Expectations.}
A notable finding from the cluster analysis is that several features, such as Chat, Voice, Scanner, Math, Answer Quality and Language, receive both positive and negative feedback. 
This duality shows the pivotal role these functionalities play in shaping user experience and reflects the high expectations users have for AI features.

For instance, the Chat functionality frequently serves as both a differentiator for user satisfaction and a source of frustration. 
On the positive side, users appreciate the quality and the chat experience, as illustrated by the following user review snippet:
\begin{userquote}
``\(\cdots\) The chats vary wildly, the responses are editable, and it helps me overcome my writer's block from time to time. \(\cdots\)''
\end{userquote}
On the contrary, other users express dissatisfaction when the chat feature demonstrates breakdowns in dialogue coherence or responsiveness:
\begin{userquote}
``\(\cdots\) The AI loops, they don't even talk about the topic, and they get uncomfy for no reason. \(\cdots\)''
\end{userquote}

This pattern is prominent across additional dual-role topics. 
For example, Answer Quality and Math both attract large volumes of positive feedback for enabling substantive, reliable assistance, yet are also among the most frequently referred to as sources of negative sentiment when answers are perceived as incomplete or erroneous. 
The existence of such polarized feedback for the same features shows the importance of not only delivering AI capabilities but ensuring their consistent and context-appropriate performance. 
In effect, these dual-role topics function as bellwethers for overall user trust and adoption: when executed well, they drive satisfaction and loyalty; when deficient, they become central pain points that can erode user confidence in the app's AI functionality.

\subsection{RQ2: Co-occurrence of Valued Features and Problems}
\begin{figure}[]
  \centering
  \includegraphics[width=\columnwidth]{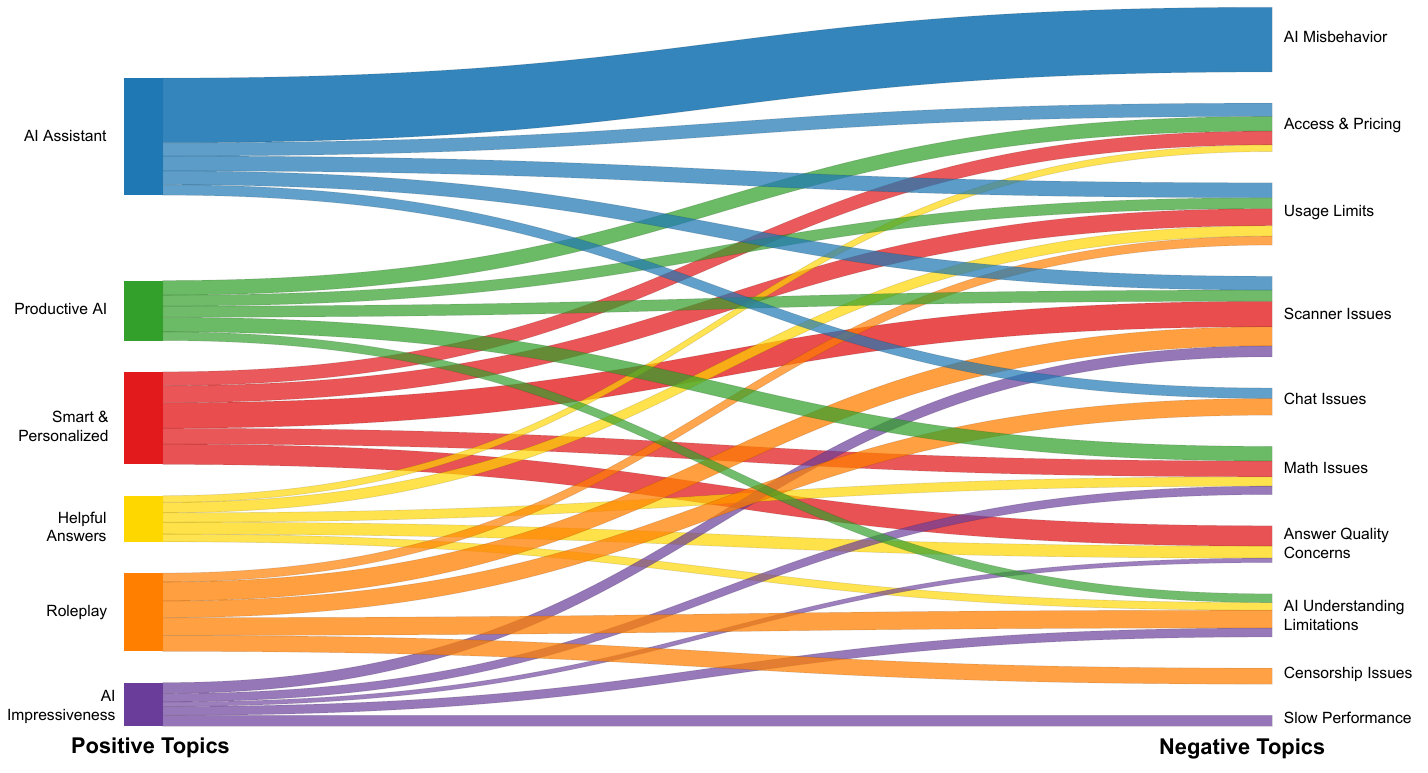}
  \caption{Visualization of the most frequent co-occurrences between leading positive features and negative problems within user reviews. Flow thickness represents the frequency of each feature–problem pairing.}
  \label{fig:co-occurance}
  \vspace{-15pt}
\end{figure}

To better understand the trade-offs users experience with AI-powered mobile apps, we analyze how positive features and negative issues co-occur within individual reviews. Fig.~\ref{fig:co-occurance} visualizes these connections, illustrating how user-valued features are often mentioned alongside reported problems. For clarity and interpretability, we filter the underlying data to include only positive topics with at least 3,000 outgoing links and, for each positive topic, the top five most frequently co-occurring negative topics. This focused view highlights the most consequential feature–problem pairings, offering insight into the complex and sometimes conflicting nature of real-world user experience.

This analysis reveals that even the most celebrated features are frequently accompanied by specific complaints. 
For instance, users are impressed by the AI's conversational depth and utility, while also noting occasional lapses in factual accuracy:
\begin{userquote} 
``Outstanding. The AI is articulate, reasonable, helpful, and intuitive. It's also got a detectably layered social aspect, as with a real person. It's not perfect – it sometimes gets well-established facts and simple math wrong, so you have to stay on your toes \(\cdots\)''
\end{userquote}

Similarly, reviews demonstrate that users praise the app's overall utility, yet express frustration with the scanner’s limitations and occasional errors: 
\begin{userquote}
``This app very useful and very good but there is many problems in this when we scan math problem it can't identify the text \(\cdots\)''
\end{userquote}

Likewise, positive feedback such as ``\(\cdots\) the personalized suggestions are great \(\cdots\)'' is often accompanied by remarks like ``\(\cdots\) the app keeps misunderstanding, \(\cdots\)'' showing how personalization sometimes falls short in execution.

We observe these patterns consistently across topics.
AI Assistant, while widely praised, is also frequently linked to user frustrations about AI Misbehavior, Access \& Pricing, and Chat Issues. 
Similarly, Smart and Personalized topics often co-occur with complaints about AI Understanding Limitations and Scanners. 
Productive AI, a popular topic, is notably paired with persistent scanner and math-related problems. 
These patterns reveal that developers and researchers should focus on addressing the specific problems most frequently paired with popular features to convert ambivalent or conflicted user feedback into strong endorsements.

\begin{figure}[]
  \centering
  \includegraphics[width=\columnwidth]{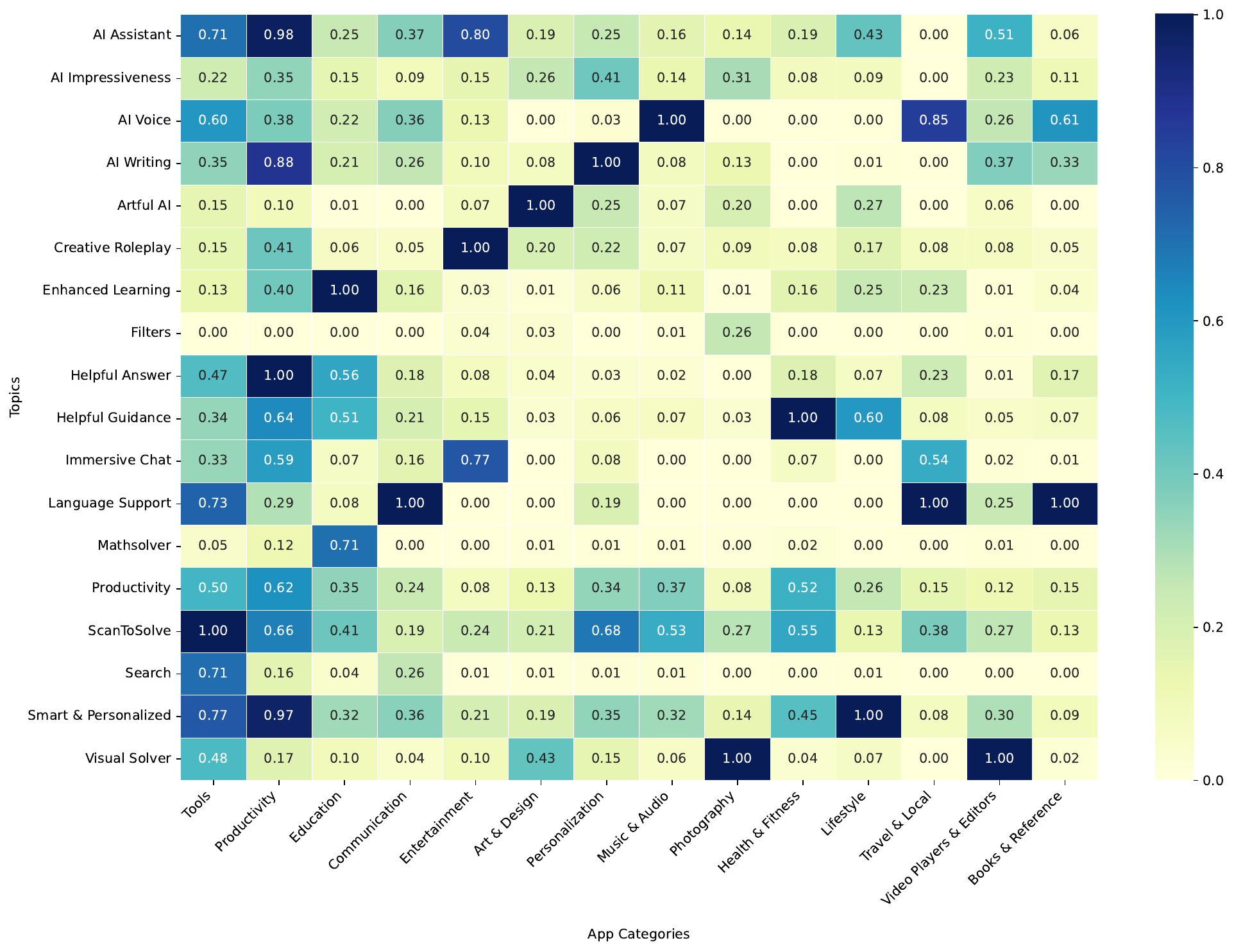}
  \caption{Normalized frequency heatmap across positive topics for each app category.}
  \label{fig:normalized_frequency_heatmap_pos-2}
  \vspace{-15pt}
\end{figure}

\subsection{RQ3: Category Contrast}
\begin{figure}[]
  \centering
  \includegraphics[width=\columnwidth]{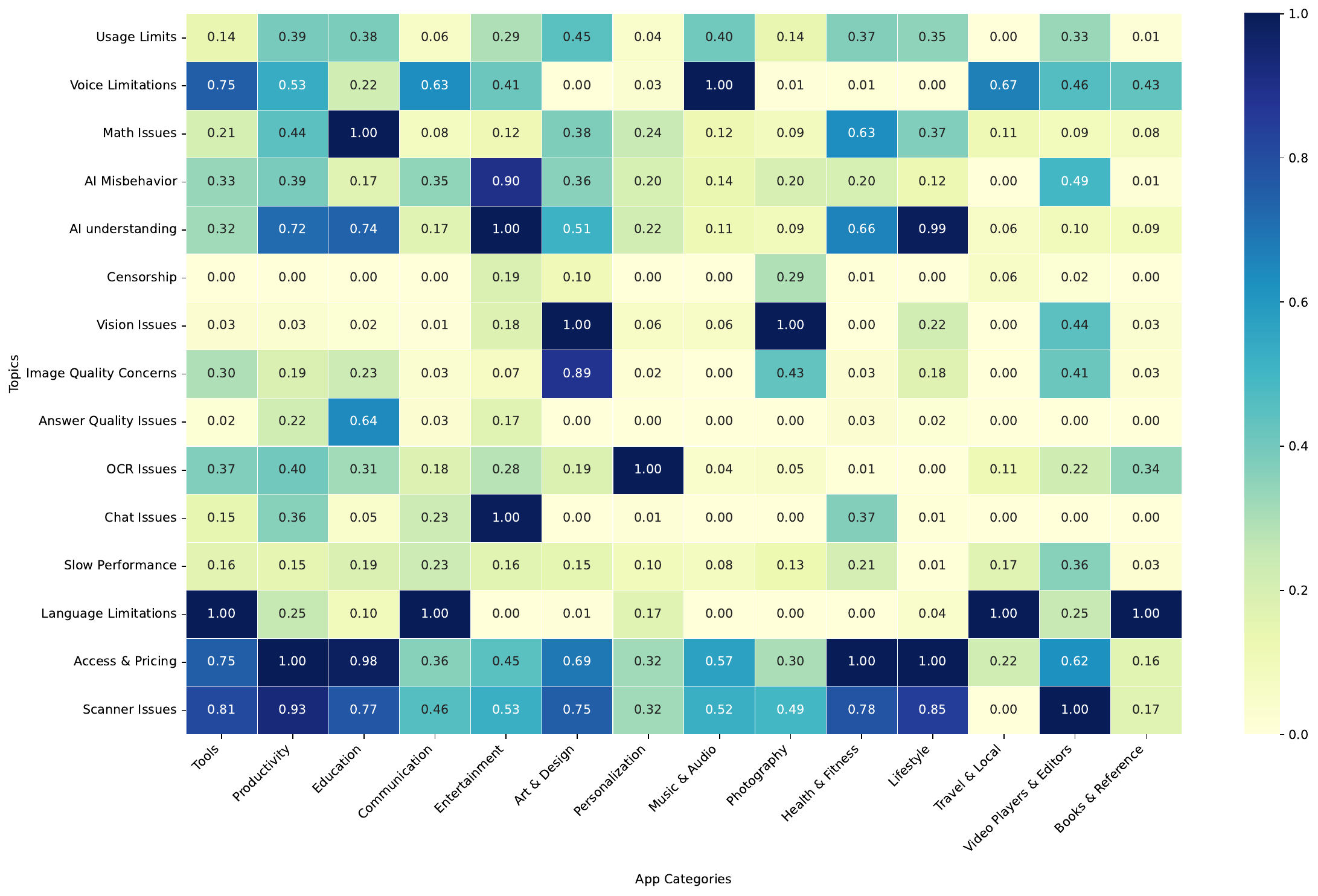}
  \caption{Normalized frequency heatmap across negative topics for each app category.}
  \label{fig:normalized_frequency_heatmap_neg-2}
  \vspace{-15pt}
\end{figure}

To analyze the distribution of user feedback topics across application domains, we constructed topic-by-category heatmaps for both positive and negative themes (see Figures~\ref{fig:normalized_frequency_heatmap_pos-2} and~\ref{fig:normalized_frequency_heatmap_neg-2}). 
To address the challenge of different category sizes, we applied min-max normalization to each column of the raw count matrix to make sure that patterns of topic salience within categories and comparisons across categories remain meaningful.

\stitle{Category-Specific Patterns.}
In Productivity apps, positive sentiment is concentrated on Helpful answers, AI Assistant, Smart \& Personalized, and AI writing, reflecting user demand for intelligent support and rapid problem-solving tools. 
In contrast, negative feedback for Productivity centers on Scanner Issues and Access \& Pricing, pointing to technical limitations in document processing and widespread dissatisfaction with paywalls and subscription models. 
Education apps show high positive feedback for Enhanced Learning and Math solver, emphasizing the value users place on academic support. 
However, negative sentiment is driven by Math Issues, Scanner Issues, and Access \& Pricing, indicating user frustration with incorrect solutions, difficulties scanning, and paywalls.

For Creative domains such as Art \& Design, users are most positive about Artful AI, which enables creative image generation and manipulation. 
Yet, the same category is marked by negative feedback around Image Quality Concerns and Vision Issues, revealing persistent challenges in the reliability of AI-powered visual features. 
In Communication, Books \& References, and Travel apps, Language Support is the leading source of positive sentiment, highlighting the growing importance of robust multilingual and conversational capabilities. 
At the same time, users in these categories also report Language Limitations as major pain points, indicating gaps in accessibility and the need for improved language. 

Overall, these category-specific patterns demonstrate that while certain features are valued universally, each domain presents its own set of high-impact technical and usability challenges that shape the user experience.

\stitle{Cross-Category Insights.}
While category-specific differences are pronounced, the results also align with prior work that a core set of features and complaints recur across multiple app domains~\cite{khalid2015what}.
AI Assistant and general-purpose assistance features are consistently valued in Productivity, Tools, and Entertainment apps, indicating a broad-based appreciation for flexible, context-aware AI. 

Features such as Smart \& Personalized are not only popular in Productivity and Tool apps but also receive positive feedback in Lifestyle apps, suggesting that problem-solving and customization are widely desired. 

On the negative side, complaints about Scanner Issues and Access \& Pricing are among the most frequent across several categories, including Tools, Productivity, Education, Health \& Fitness, and Lifestyle. 
This finding suggests that technical reliability and fair monetization remain persistent, unresolved issues for AI-powered mobile apps. 
Language and capabilities also emerge as a cross-cutting concern: both positive (Language Support) and negative (Language Limitations) topics are highly prevalent in Tools, Communication, Books \& References, and Travel apps. 
These cross-category results reinforce the idea that, while users in different domains may emphasize different features, a handful of technical and business model factors dominate the user experience landscape for AI-driven mobile apps. 
Addressing these systemic issues—such as improving scanner reliability, reducing monetization friction, and strengthening multilingual support would yield broad improvements across diverse app categories. 
For developers and researchers, these findings highlight the importance of targeting both universal and category-specific challenges to optimize the impact of AI features in mobile software.

\section{Implications}
Drawing on the empirical findings of RQ1 -- RQ3, we derive a set of actionable recommendations that address both overall and category-specific user pain points.

\begin{enumerate}[label=\emph{(\alph*)}]
    \item \emph{Strengthen reliability in core AI features.} 
    Failures in scanning, OCR, and vision collectively represent some of the most common sources of user frustration across multiple categories. Scanner-related issues account for over 13.2K negative aspects and are among the leading complaints in Tools, Productivity, Education, Health \& Fitness, and Lifestyle apps. To address this, developers should prioritize rigorous validation, continuous monitoring, and user-centered error handling for scanning features.
    
    \item \emph{Revisit monetization and access models.}
    Negative sentiments concerning pricing, usage limits, and paywalls were observed in over 17.7K aspect mentions, with monetization complaints constituting one of the most prominent issues across Tools, Productivity, Education, Health \& Fitness, and Lifestyle apps.
    Developers should consider transparent pricing, generous trial options, and flexible tiered plans -- especially in productivity and education domains, where access barriers most acutely undermine perceived value.

    \item \emph{Addressing Valued Feature-Linked User Frustrations.} To maximize positive impact, developers should prioritize addressing the specific problems most commonly linked to their app's most liked features. For example, if a math-solving feature receives positive feedback but also draws complaints about occasional errors, prioritizing solutions around highlighted errors can boost user satisfaction.

\item \emph{Addressing Dual-Role Topics.}
    Certain topics, such as language support, receive both praise and criticism from users, highlighting their importance for overall satisfaction. For instance, while multilingual capabilities are often celebrated, any gaps in localization are quickly noted as major drawbacks. Developers should focus on improving consistency in these dual-role areas to convert mixed feedback into lasting strengths.
\end{enumerate}

\stitle{Category-Specific Recommendations:}
\begin{itemize}
    \item \emph{Productivity, Tools \& Education:} 
    Developers should focus on enhancing the accuracy and reliability of problem-solving features such as scanning, OCR, and math solvers, as these are frequent sources of both user praise and complaint. 

    \item Communication, Travel \& Local and Books \& Reference : Invest in multilingual support and customizable voice and language options, as missing support is a leading cause of user frustration.
    \item \emph{Art \& Design, Photography and Entertainment:} Prioritize creative freedom, style diversity, and expressive roleplay. Negative reviews often cite visual recognition flaws, ineffectiveness, and lack of understanding, while positive reviews celebrate creative output.
\end{itemize}

\stitle{Summary.} 
The strongest opportunities for improving AI-powered mobile apps are concentrated around a handful of core themes: technical reliability, pricing, language inclusivity, expressive features, and responsive moderation. 
The data-driven analysis in this study provides clear guidance on which aspects most demand developer attention in each app category and highlights the value of ongoing, systematic user feedback analysis for sustaining competitive, user-aligned AI products.

\section{Threats to Validity}
\label{threats}
\stitle{Construct Validity.}
Our definition of ``AI-related'' reviews and aspect extraction depends on annotation guidelines and prompt design. 
While validated, some ambiguity and misclassification may persist, especially for implicit or nuanced mentions of AI.

\stitle{Internal Validity.}
Model selection, prompting, and clustering choices may influence the results. 
Despite benchmarking and human validation, LLMs limitations and occasional subjective judgments could introduce bias.

\stitle{External Validity.}
Our findings are based on Google Play Store data for 292 apps and may not generalize to other platforms, app stores, or user populations. 
Non-English reviews and niche domains are underrepresented.

\stitle{Reliability.}
While all data, code, and prompts are released for replication, results may vary as LLMs, app content, and reviews evolve.

\stitle{Mitigation.}
We mitigate these threats by making all data, code, prompts, and analysis procedures openly available, enabling transparent replication and critical assessment by the community.

\section{Related Work}
\label{sec:related}
\stitle{User Review Mining in Mobile Apps.}
Prior work established the feasibility of mining large-scale reviews to identify usability and privacy issues~\cite{privacy-at-scale}. 
Unsupervised and hybrid methods have since been introduced for discovering privacy concerns, using domain knowledge and embedding-based semantic filtering to surface issues often overlooked by supervised or topic modeling techniques~\cite{Ebrahimi,hybrid,glove}.
Accessibility feedback has also been mined, using automated keyword detection and qualitative review analysis to inform inclusive app design~\cite{Haystack,Reyes}. 
Recent works combine deep learning and NLP to classify millions of reviews and generate high-level summaries of privacy or usability themes over time~\cite{Large-Scale-Trends-Privacy,hark}.
While this research have improved the breadth and efficiency of review mining, the literature still lacks systematic, large-scale studies of user perspectives on AI-powered features across the increasingly diverse landscape of mobile apps.

\stitle{Aspect Extraction and Topic Modeling.}
Fine-grained analysis of user feedback often relies on aspect-based sentiment analysis (ABSA), which has evolved from aspect extraction and sentiment classification to more complex tasks such as end-to-end ABSA, triplet extraction, and aspect-opinion pairing~\cite{ABSA-SURVEY,ate,asc,ACD,OTE,e2e,aste,aope,asqp}.
These works enable the detection of user sentiment at the feature or aspect level. 
For higher-level themes, topic modeling methods like LDA and NMF~\cite{LDA,nmf} have long been used, but often falter on short, informal review texts. Recently, transformer-based methods such as BERTopic~\cite{bertopic} have improved topic coherence by leveraging contextual embeddings.
Building on these works, our study adopts a unified approach: extracting AI-specific aspect–sentiment pairs using LLM-based embeddings, then applying K-Means clustering to identify coherent topics and actionable themes in user feedback.

\stitle{AI-Specific Concerns in App Reviews.}
With the spread of AI-based features, several studies have targeted fairness, privacy, and security in AI-based mobile apps. 
For instance, recent research analyzes fairness in user experiences with AI apps by mining and clustering relevant reviews~\cite{Fairness}. 
Other work~\cite{empirical-ai} explores AI-enabled apps from technical and user-centered perspectives, filtering for AI-related feedback using technical dictionaries and sentiment analysis. 
Security issues in LLM-powered apps have also been addressed using automated frameworks for review analysis~\cite{llmsecurity}.
However, these studies often depend on explicit keyword expansion or manual review, which is not scalable given the the volume, variability, and unstructured nature of user feedback.
To this end, our work offers a scalable solution by leveraging LLM-driven semantic classification and clustering, enabling sentiment- and category-aware mapping of both explicit and implicit user experiences with AI-powered app features.

\section{Conclusion and Future Work}
This study presents the first comprehensive, large-scale analysis of user feedback on AI-powered features in mobile apps. Using a curated dataset of 894K AI-specific reviews from 292 apps across 14 categories, we develop and validate an automated pipeline that combines large language models with semantic clustering to extract fine-grained, interpretable insights at scale. 
Our findings show that while users consistently praise productivity, intelligent assistance, and personalization, they also express frustration over technical failures, pricing barriers, and limited language support. 
Importantly, our method captures both positive and negative feedback within the same review, revealing how users often appreciate a feature's utility while simultaneously expressing frustration with its limitations. This contrast is often missed by analyses that treat sentiment in isolation, resulting in an incomplete view of user experience.
To make sense of the extracted aspect–sentiment pairs at scale, we cluster them into 18 positive and 15 negative themes based on semantic similarity. Our analysis highlights how user satisfaction and frustration with AI features vary both broadly and within specific app categories.
Our publicly released data, pipeline code, and analysis artifacts can provide a foundation for future research on user-centered AI-based feature design in mobile apps.
Looking ahead, extending this work to non-English reviews, additional app platforms, and longitudinal trends will yield deeper insights into evolving user expectations. 

\balance
\bibliographystyle{plain}
\bibliography{sample-base}

\end{document}